\documentstyle[preprint,aps]{revtex}
\begin{document}
\title{The Role of Mass and External Field on the Fermionic Casimir Effect}
\author{J. L. Tomazelli} 
\address{Departamento de F\'{\i}sica e Qu\'{\i}mica, Faculdade de Engenharia,
Universidade Estadual Paulista, Campus da Guaratinguet\'a,
Av. Dr. Ariberto Pereira da Cunha 333, 
12500-000 \\ Guaratinguet\'a, SP, Brazil.}
\author{L. C. Costa}
\address{Instituto de F\'{\i}sica Te\'{o}rica,\\
Universidade Estadual Paulista, \\
01405-900, S\~{a}o Paulo, Brazil.}
\maketitle
{\small{
\begin{abstract}
The aim of this work is to investigate the role played by the fermion mass and that of an external 
field on the fermionic Casimir energy density under $S^1 \times R^3$ topology. Both twisted and 
untwisted spin connections are considered and the {\it exact} calculation is performed using a 
somewhat different approach based on the combination of the analytic regularization method through 
$\alpha$-representation and the Euler-Maclaurin summation formula.
\end{abstract}
%
\renewcommand{\thefootnote}{\fnsymbol{footnote}} 
\def\0{\begin{equation}}
\def\1{\end{equation}}
\def\2{\begin{eqnarray}}
\def\3{\end{eqnarray}}
\def\>{{\rangle}}
\def\<{{\langle}}
\def\H{{\rm {H}}}
\def\p{{\bf {p}}}

%
%
%
\newpage
\section{Introduction}
As well as the Lamb-Retherford shift of atomic energy levels and the electron's anomalous magnetic 
moment, the Casimir force between two parallel perfectly conducting plates is one of the most remarkable 
manifastations of quantum vacuum fluctuantions. It was first predicted on theoretical grounds by H. B. G. 
Casimir in 1948 \cite{CA48} and experimentally verified on a qualitative level by Sparnaay \cite{SP58} 
ten years later. Recentely, high accuracy experiments have been performed by Lamoreaux \cite{LA97} and by 
Mohideen and Roy \cite{MO98}. For a more detailed account on the subject there are excellent reviews in 
the literature \cite{PL86}.

In 1975, employing Casimir essential ideas, Johnson \cite{JO75} investigated the effects of boundaries 
on a massless Dirac field in the context of MIT-bag model \cite{CH74} and found an energy density 
shift of the same order of magnitude as that obtained by Casimir for the electromagnetic field. Addopting 
Johnson's extend approach to Casimir effect, in order to allow for other quantum fields, many authors have 
investigated the effects of different boundary conditions on the fields considered \cite{DE79}. Hence, one 
can say that a modern view of Casimir effect might take into account those effects caused by non-trivial 
space topologies on the vacuum of quantum fields. 

It is worthnoting that in this general context some Casimir setups (field + boundary condition 
+ external sources) present quite complicated final expressions for the Casimir energy density, 
which ultimately obscure any possible physical interpretation. In particular, the results obtained 
in \cite{FO80} for the massive spinor field indicate a mass dependent energy density which calls 
for a deep investigation. Similar difficulties also appear in the case considered in \cite{CO01} 
where the fermion field is also subjected to an external magnetic field. 

In order to handle the above mentioned shortcomings, we present here an alternative treatment which 
allows us to extract new information concerning the role of mass and that of an external field to the 
fermionic Casimir effect. This is achived by a suitable combination of the method of analytic 
regularization in the context of gamma function representation (also called $\alpha$-representation 
\cite{BO59}) and the use of the well-known Euler-Mclaurin summation formula \cite{LE82}. 

The main ideas of our construct are introduced in the next section, where we consider the 
well-stablished electromagnetic Casimir effect. In section III, the same procedure is employed 
to the case of a massive spinor field with boundaries analogous to that considered in \cite{DE79} 
and \cite{FO80}. In section IV, the effects of an external constant and homogeneous magnetic field 
is also considered and the connection with the so-called Euler-Kockel-Heisenberg Effective Lagrangian 
density \cite{EU36} is addressed. Finaly, in section V, we make some concluding remarks pointing out 
directions of future investigations.   
%
%
\section{The Case of Electromagnetic Fields}
As was originally considered by Casimir in 1948, the divergent vacuum energy density associated 
with the electromagnetic field is given by 
\0
\varepsilon_0 = \frac{1}{8 \pi^2 L} \sum_{n= - \infty}^{+ \infty} 
\int_{- \infty}^{\infty} \int_{- \infty}^{\infty} dk_x \; dk_y \; \frac{1}{[ k_x^2 + k_y^2 + (an)^2 ]^{-1/2}}
\;,\;\;\;\;\;a = \pi/L
\1
where the discrete values of $k_z$ reflects the boundary condition imposed by two parallel conducting plates 
with area $A = l^2$ and separated by a distance $L$ in such a way that $l >> L$ ($a = \pi/L$) \cite{CA48}. 

In (1), the integrals are quadraticaly divegent quantities which claim for a consistent 
regulariazation prescription. Despite the familiar regularization methods found in the 
literature \cite{BO59} \cite{BO90}, we shall consider here a quite different one. It consists 
in the conbination of the analytic regulariazation scheme, using the gamma function integral 
representation, with the so-called Euler-Mclaurin summation formula \cite{LE82}. To see how 
this works, we start by taking the analytic extension of the integrand in (1), which turns 
out to be a regular functional. This is achieved by means of the gamma function integral 
representation 
\0
\frac{1}{A^{1 + \delta}} = \frac{1}{\Gamma (1 + \delta)} \int_{0^+}^{\infty}
d\eta \; \eta^{\delta} {\rm e}^{- A \eta},
\1
valid for $\delta > - 1$, which allows us to rewrite (1) as
\0
(\varepsilon_0)^{\rm R} = \frac{1}{2 (2 \pi)^2 L} \frac{\pi}{\Gamma(- 1/2 + \delta)} \sum_{n= - \infty }^{\infty} 
\int_{0^+}^{\infty} d\eta \; \eta^{- 5/2 + \delta} {\rm e}^{-(an)^2 \eta}, 
\1 
where the gaussian integrals in $k_x$ and $k_y$ have already been calculated. Note that, for 
$\delta \rightarrow 0$, we expect to recover the original theory, i.e., expression (1). This will 
be done only at the end of the calculations.  

The divergent sum over $n$ appearing in (3) is performed by means of the Euler-Mclaurin summation formula 
\cite{LE82}
\2
\sum_{n = M}^{N} f(n) &=& \int_M^N F(x) dx + \frac{1}{2} [ f(N) + f(M) ] + 
\sum_{k=1}^{K} \frac{B_{2k}}{2k!} [F^{2k -1}(N) - F^{2k -1} (M) ] \nonumber \\
& & + \frac{1}{(2K + 1)!} \int_N^M B_{2K +1} (x - [x]) F^{2K + 1} (x) dx
\3
where $B_m \equiv B_m(0)$ and the $B_m (x)$ are the Bernoulli polynomyals. We preserve here the same 
notation used in \cite{LE82}. The last term in (4), also called the remainder term, vanishes if $F(z)$ 
is an entire function. Further, if $n$ is integer and $M \leq n \leq N$, then $F(n) = f(n)$. In the 
present context, identifying the entire function $f(n)$ with 
\0
f(n) = {\rm e}^{- (an)^2 \eta}
\1
and, since $0 \leq n \leq \infty$, we are allowed to rewrite (3) as
\2
(\varepsilon_0)^{\rm R} &=& \frac{1}{(2 \pi)^2 L} \frac{\pi}{\Gamma(- 1/2 + \delta)} \left\{ 
\int_{0^+}^{\infty} d\eta \;  \eta^{- 5/2 + \delta} \left[ 
\int_{0}^{\infty} dn \; f(n) - \frac{1}{2} f(n)|_{n \rightarrow \infty} 
\right. \right. \nonumber \\ 
& & \left. \left. + \frac{1}{12} f'(n)|_{n \rightarrow \infty} - \frac{1}{12} f'(n)|_{n \rightarrow 0} - 
\frac{1}{720} f'''(n)|_{n \rightarrow \infty} + \frac{1}{720} f'''(n)|_{n \rightarrow 0} + ...  
\right] \right\} 
\3
where $f'(n)$ means the first derivative of (5) with respect to $n$ and so on. Calculating the 
derivatives and taking the corresponding limits we see that the only non-vanishing terms give
\2
(\varepsilon_0)^{\rm R} &=& \frac{1}{(2 \pi)^2 L} \frac{\pi}{\Gamma(- 1/2 + \delta)} \left\{ 
\int_{0^+}^{\infty} d\eta \;  \eta^{- 3 + \delta} \frac{\sqrt{\pi}}{2 a }
+ \frac{1}{12} \left( \frac{\Gamma (-1/2 + \delta)}{(a^2n^2)^{-1/2 + \delta}} 
(-2a^2n) \right)_{n \rightarrow \infty} \right. \nonumber \\
& & \left. - \frac{1}{720} \left( \frac{\Gamma (3/2 + \delta)}{(a^2n^2)^{3/2 + \delta}} 
(-2a^2n)^3 \right)_{n \rightarrow \infty} - 
\frac{3}{720} \left( \frac{\Gamma (1/2 + \delta)}{(a^2n^2)^{1/2 + \delta}} 
(-2a^2)^2 n \right)_{n \rightarrow \infty}  \right\}
\3
where (2) was used. In going from expression (6) to (7) we notice that only the ($n \rightarrow \infty$)-
terms contribute to the energy density. Usually, the methods found in the literature extract information 
arising from the ($n \rightarrow 0$)-terms and, as we will show in the next section, this generates quite 
different final results. However, we must stress that, while the exponential function in (5) is an 
analytical function over the {\em entire} complex plane, the power function in the integrand of (1) is a 
multiple-valued function, which has a branch cut along the real axis \cite{CO74}. 

Now, in order to obtain a consistent result with the original theory we now take the limit 
$\delta \rightarrow 0$ in (7). First, we must handle the divergent contribution arising from the 
second term in the curly brackets. Such divergence is eliminated using the freedom in the choice of 
$\delta$ in (7). In fact, for consistency with (3), $\delta$ are constrained to be greater than $1/2$. 
However, to obtain the correct final result the considered region in the complex plane must be 
analytic continued to $\delta \geq 1$. 
Hence, as the limit $n \rightarrow \infty$ is performed, the second term contribution turns out to 
be zero. On the other hand, the contribution from the remaining terms, which becomes $n$-independent 
when the limit $\delta \rightarrow 0$ is taken, gives
\0
\varepsilon_0 = - \frac{1}{4(2 \pi)^2} \int_0^{\infty} d \eta 
\; \eta^{- 3} + \frac{\pi^2}{720 L^4}.
\1
Finally, calculating the corresponding energy for the whole space (which is equivalent to 
taking $L \rightarrow \infty$ in the above expression) and subtracting from it expression 
(8), yields 
\2
\Delta \varepsilon &=& ( \varepsilon_0 )_{L \rightarrow \infty} - \varepsilon_0  \nonumber \\
&=& - \frac{\pi^2}{720 L^4}
\3
which is the expected vacuum energy density derived by Casimir in 1948. Since there is no {\it a priori} 
reason for assuming that the vacuum energy in the presence of boundaries is greater than in their absense, 
we have defined $\Delta \varepsilon$ in such a manner that (9) indicates that the force between the 
plates is in fact attractive \footnote{Once we adopt this definition, it must be preserved if we 
consider other topologies and boundary conditions, in order to be able to distinguish attractive from 
repulsive effects.}, as has been confirmed by precise experiments \cite{LA97} - \cite{MO98}.
%
%
\section{The Massive Spinor field: $S^1 \times R^3$}
The case of noninteracting spinor fields subjected to $S^1 \times R^3$ space topology with twisted and 
untwisted spin connections was first considered by DeWitt, Hart and Ishan in 1979 \cite{DE79} and one 
year later it was generalized to the massive case by Ford \cite{FO80}, who obtained an intrincate mass 
dependent expression for the vacuun energy density. In this section we intend to rederive these results 
using the procedure developed in the last section. As will be shown, the effect of the fermion mass to 
the Casimir energy density is, in fact, null. 
%
%
\subsection{Untwisted Case}
For an untwisted spinor field, the vacuun energy density is  given by 
\0
\varepsilon_0^{\rm unt} = - \frac{1}{(2 \pi)^2 L} \sum_{- \infty}^{\infty} \; 
\int_{- \infty}^{\infty} dp_x \; dp_y \; [ m^2 + p_x^2 + p_y^2 + p_z^2]^{1/2}.
\1
where $p_z = a^2 n^2$, with $n = 0, \pm 1, \pm 2, ...$ and $a = 2 \pi/L$. As in the electromagnetic 
case, the above quantity is divergent and must also be regularized. Using the same procedure employed 
in the last section, we write (10) as
\2
(\varepsilon_0^{\rm unt})^R &=& - \frac{1}{2 \pi L} \frac{1}{\Gamma(- 1/2 + \delta)} \sum_{n= - \infty}^{\infty} 
\int_0^{\infty} \; d\eta \; \eta^{- 5/2 + \delta} \; {\rm e}^{- (m^2 + a^2 n^2) \eta} \nonumber \\
&=& - \frac{1}{2 \pi L} \frac{1}{\Gamma(- 1/2 + \delta)} 
\int_0^{\infty} \; d\eta \; \eta^{- 5/2 + \delta} \left[ {\rm e}^{- m^2 \eta} + 
2 \sum_{n = 1}^{\infty} {\rm e}^{- (m^2 + a^2 n^2) \eta} \right], 
\3
where we have already performed two gaussian integrals in $p_x$ and $p_y$. Instead of use the 
Abel-Plana formula \cite{ER53}-\cite{MO88} we may invoke the Euler-Mclaurin summation formula 
(4), in order to find an expression similar to (6), where the only difference rely upon the 
definition of the entire function $f(n)$. Here
\0
f(n) = {\rm e}^{- (m^2 + a^2 n^2) \eta}.
\1
Following the same steps done in the last section, it is straightforward to obtain
{\footnotesize{
\2
(\varepsilon_0^{\rm unt})^R &=& - \frac{1}{(2 \pi) L} \frac{2}{\Gamma(- 1/2 + \delta)} \left\{ 
\int_{0^+}^{\infty} d\eta \;  \eta^{- 3 + \delta} {\rm e}^{- m^2 \eta}\frac{\sqrt{\pi}}{2 a} 
+ \frac{1}{12} \left( \frac{\Gamma (-1/2 + \delta)}{(m^2 + a^2n^2)^{-1/2 + \delta}} 
(-2a^2n) \right)_{n \rightarrow \infty} \right. \nonumber \\
&-& \left. \frac{1}{720} \left( \frac{\Gamma (3/2 + \delta)}{(m^2 + a^2n^2)^{3/2 + \delta}} 
(-2a^2n)^3 \right)_{n \rightarrow \infty}
- \frac{3}{720} \left( \frac{\Gamma (1/2 + \delta)}{(m^2 + a^2n^2)^{1/2 + \delta}} 
(-2a^2)^2 n \right)_{n \rightarrow \infty}  \right\}. 
\3}}
Analyzing the structure of (13), we see that the fermion mass appearing in the denominator of the 
last three terms may be neglected since in those terms $n \rightarrow \infty$. This point is crucial 
since the limit accounts for the partial elimination of the fermion mass. In fact, $m$ remains in the 
kernel of the first term but, as will shown later, this term will be cancelled against the Minkowiski 
vacuum energy . The divergence of the second term in (7) also appears and we should say that it is a common 
feature of our approach. Again, it is eliminated by taking an analytic extension analogous to that 
employed in the electromagnetic case. As a final result we have
\0
\varepsilon_0^{\rm unt} = \frac{1}{8 \pi^2} \int_0^{\infty} d\eta \; 
\eta^{-3} {\rm e}^{- m^2 \eta} - \frac{\pi^2}{720 L^4}.
\1
Subtracting (14) from the corresponding usual Minkowiski vacuun energy (which correspond to taking 
$L \rightarrow \infty$ in (14)) we find the fermionic Casimir energy density 
\0
\Delta \varepsilon^{\rm unt} = \frac{2 \pi^2}{45 L^4},
\1
which is in complete agreement with \cite{DE79}, the only difference being a factor $2$, which reflects 
the four-component spinor field representation we are using. De Witt {\it et al} considered a two-component 
spinor field. 
%
%
\subsection{Twisted Case}
Expression (14) clearly shows the independence of the Casimir energy density with respect to the 
fermion mass. This feature also occurs when twisted boundary condition is considered. In this case 
$p_z = (2n+1) a$, with $n=0, \pm 1, \pm 2, ...$ and $a = \pi/L$. The regulated vacuun energy density 
is now given by  
\0
(\varepsilon_0^{\rm twi})^R = \frac{\sqrt{\pi}}{(2 \pi)^2 L} \sum_{n = - \infty}^{\infty} \int_0^{\infty} 
d \eta \; \eta^{- 5/2} \; {\rm e}^{- (m^2 + (2n+1)^2 a^2) \eta}.
\1
Since the above integral is regular, we are allowed to interchange the sum and the integral, and than 
use the following mathematical trick
\0
\sum_{n = - \infty}^{\infty} {\rm e}^{- (m^2 + (2n+1)^2 a^2) \eta} = 
\sum_{n = - \infty}^{\infty} {\rm e}^{- (m^2 + n^2 a^2) \eta} - 
\sum_{n = - \infty}^{\infty} {\rm e}^{- (m^2 + (2n)^2 a^2) \eta}. 
\1
In this way the problem of solving the twisted case reduces to that of computing two terms proportional 
to that in the untwisted case. In fact, we have
\2
\varepsilon_0^{\rm twi} &=& \frac{1}{2^3} \varepsilon_0^{\rm unt} - \varepsilon_0^{\rm unt} \nonumber \\
&=& - \frac{7}{8 (2 \pi)^2} \int_0^{\infty} \; d \eta \; \eta^{- 3} + 2 \frac{7 \pi^2}{360 L^4}.
\3
Again, subtracting this result from that where $L \rightarrow \infty$, we obtain
\0
\Delta \varepsilon_0^{\rm twi} = - 2 \frac{7 \pi^2}{360 L^4},
\1
which coincide with the familiar result found in the literature for the massless fermionic Casimir effect 
\cite{DE79}.
%
%
\section{The Influence of an External Field}
Another interesting problem to be analysed using the present construct is related to influence of 
an external magnetic field on the Casimir energy associated to the Dirac field. This problem has 
been recentely proposed in the context of Effective Quantum Electrodynamics using the so-called 
Schwinger proper-time method \cite{CO01}. However, a clear answser concerning the role of the external 
field on the fermionic Casimir energy density is yet an open problem which deserves further investigation. 
The purpose of this section is to implement, in the context of Weisskopf method \cite{WE36} \cite{JE00}, 
the prescription presented in the previous sections in order to get a better understanding of the above 
mentioned problem.

We restrict our calculation to the case where an untwisted massless spinor field is subjected to 
an external constant uniform magnetic field. As is well known \cite{BE97}, the negative energy levels 
for an electron of charge $e = - \vert e \vert$ in the presence of an uniform and constant magnetic 
field $\H_z = - \H$ is given by 
\0
- \epsilon_{\p, \sigma}^{(-)} = - \sqrt{ m^2 + (2n + 1 - \sigma) \vert e \vert \H + p_z^2 },
\1
where $ n = 0, 1, 2, 3 ...$ and $ \sigma = \pm 1 $. Taking into account the density of states in the 
interval $dp_z$
\0
\frac {\vert e \vert \H}{2 \pi} \frac{dp_z}{2 \pi} \nonumber 
\1
and the fact that all the levels except $n=0, \sigma = 1$ are doubly degenerate (the levels 
$n, \sigma = -1$ and $n+1, \sigma = 1$ coincide), we obtain the energy density of vacuum 
electrons,
\2
\varepsilon_0' &=& - \sum_{\p, \sigma} \epsilon_{\p, \sigma}^{(-)} \nonumber \\
&=& - \frac{\vert e \vert \H}{(2 \pi)^2} \int_{- \infty}^{+ \infty} \left\{ 
\sqrt{ p_z^2 } + 2 \sum_{n=1}^{\infty} \sqrt{2 \vert e \vert \H n + p_z^2 } \right\} 
dp_z.
\3
where $p_z$ turned out to be a discrete quantity in virtue of the untwisted $S^1 \times R^3$ space topology 
we are assuming. Using (2), the energy density (22) may be rewritten in the more convenient form,
\2
(\varepsilon_0')^R &=& - \frac{\vert e \vert H}{2 \pi L} \sum_{n = 0}^{\infty} \frac{1}{\Gamma(- 1/2 + \delta)} 
\int_{- \infty}^{\infty} d\eta \; \eta^{-3/2 + \delta} \left( \sum_{n'= - \infty}^{\infty} 
{\rm e}^{-(2 \vert e \vert \H n + a^2 n'^2) \eta} \right) \nonumber \\
&=&  - \frac{\vert e \vert H}{2 \pi L} \sum_{n = 0}^{\infty} \frac{1}{\Gamma(- 1/2 + \delta)} 
\int_{- \infty}^{\infty} d\eta \; \eta^{-3/2 + \delta} {\rm e}^{- \alpha \eta} 
\left( f(0) + 2 \sum_{n'=1}^{\infty} f(n') \right) 
\3
with $2 \vert e \vert \H n = \alpha (n) \equiv \alpha$,  
\0
f(n') = {\rm e}^{- (a n')^2) \eta},
\1
and $n' = 0, \pm 1, \pm 2, ...$, $a = 2 \pi/L$. Aplying the Euler-Maclaurin formula (4) and performing 
the corresponding derivatives and limits, we arrive at 
{\small{
\2
(\varepsilon_0')^R &=& - \frac{\sqrt{\pi}}{(2 \pi)^2} \frac{\vert e \vert \H}{\Gamma(-1/2 + \delta)} 
\int_0^{\infty} d\eta \; \eta^{-2 + \delta} \; \sum_{n=0}^{\infty} {\rm e}^{- \alpha \eta} \nonumber \\
& & \frac{2}{3 L^3} \frac{\vert e \vert \H}{\Gamma(-1/2 + \delta)} 
\left\{ n' \int_0^{\infty} d\eta \; \eta^{- 1/2 + \delta} \; {\rm e}^{- (a n)^2 \eta}
\sum_{n=0}^{\infty} {\rm e}^{- \alpha \eta} \right\}_{n' \rightarrow \infty} \nonumber \\
& & - \frac{2^4 (2 \pi)^5}{720 L^7} \frac{\vert e \vert \H}{\Gamma(-1/2 + \delta)} 
\left\{ n'^3 \int_0^{\infty} d\eta \; \eta^{3/2 + \delta} \; {\rm e}^{- (a n)^2 \eta}
\sum_{n=0}^{\infty} {\rm e}^{- \alpha \eta} \right\}_{n' \rightarrow \infty} \nonumber \\
& & + \frac{12 (2 \pi)^3}{720 L^5} \frac{\vert e \vert \H}{\Gamma(-1/2 + \delta)} 
\left\{ n' \int_0^{\infty} d\eta \; \eta^{1/2 + \delta} \; {\rm e}^{- (a n)^2 \eta}
\sum_{n=0}^{\infty} {\rm e}^{- \alpha \eta} \right\}_{n' \rightarrow \infty}.
\3}}
The sum in the integrands can be eliminated by noting that 
\0
\sum_{n=0}^{\infty} {\rm e}^{- \alpha \eta} =
\sum_{n=0}^{\infty} {\rm e}^{- 2 \vert e \vert \H \eta} = \coth(\vert e \vert \H \eta).
\1
Furthermore, assuming the weak field regime ($\H << 1$) we are allowed to expand the kernel of 
the integrals in (25), namely,
{\small{
\2
(\varepsilon_0')^R &=& - \frac{\sqrt{\pi}}{(2 \pi)^2} \frac{\vert e \vert \H}{\Gamma(-1/2 + \delta)} 
\int_0^{\infty} d\eta \; \eta^{-2 + \delta} \; \coth(\vert e \vert \H \eta) \nonumber \\
& & \frac{2}{3 L^3} \frac{\vert e \vert \H}{\Gamma(-1/2 + \delta)} 
\left\{ n' \int_0^{\infty} d\eta \; \eta^{- 1/2 + \delta} \; {\rm e}^{- (a n)^2 \eta}
\left( \frac{1}{\vert e \vert \H \eta} + \frac{\vert e \vert \H \eta}{3} + \Sigma
\right) \right\}_{n' \rightarrow \infty} \nonumber \\
& & - \frac{2^4 (2 \pi)^5}{720 L^7} \frac{\vert e \vert \H}{\Gamma(-1/2 + \delta)} 
\left\{ n'^3 \int_0^{\infty} d\eta \; \eta^{3/2 + \delta} \; {\rm e}^{- (a n)^2 \eta}
\left( \frac{1}{\vert e \vert \H \eta} + \frac{\vert e \vert \H \eta}{3} + \Sigma
\right)   \right\}_{n' \rightarrow \infty} \nonumber \\
& &  \frac{12 (2 \pi)^3}{720 L^5} \frac{\vert e \vert \H}{\Gamma(-1/2 + \delta)} 
\left\{ n' \int_0^{\infty} d\eta \; \eta^{1/2 + \delta} \; {\rm e}^{- (a n)^2 \eta}
\left( \frac{1}{\vert e \vert \H \eta} + \frac{\vert e \vert \H \eta}{3} + \Sigma
\right)  \right\}_{n' \rightarrow \infty}.
\3}}
where 
\0
\Sigma \equiv \sum_{k=2}^{\infty} \frac{2^{2k} B_k}{(2 k)!} (\vert e \vert \H \eta)^{2k - 1},
\1
and the $B_k$'s are the Bernoulli numbers. 

We are now in position to perform, term by term in the expansion of integral (27), the limit 
$n' \rightarrow \infty$. After a straightforward calculation we obtain
\0
\varepsilon_0' = \frac{\vert e \vert \H}{8 \pi^2} \int_0^{\infty} d\eta \; \eta^{-2} 
\coth(\vert e \vert \H \eta) - \frac{2 \pi^2}{45 L^4},
\1
where the same kind of analytic extension made in the previous sections was performed in the 
manipulation of the first term in the second line of (27). Again, it gives no contribution. 

Finaly, the energy density of the empty space may be obtained by taking the limit of zero field 
and infinite volume in (29). We must subtract (29) from this quantity, obtaining
\0
\Delta \varepsilon_0 = - \frac{1}{8 \pi^2} \int_{0^+}^{\infty} \frac{d \eta}{\eta^3} \left\{ 
\vert e \vert H \eta \coth (\vert e \vert H \eta) - 1 \right\} + \frac{2 \pi^2}{45 L^4},
\1
which clearly shows the influence of the external magnetic field to the fermionic Casimir effect. 
It must be noted that the above expression recovers (15) in the limit of zero magnetic field. 
In addition, the first term in (30) might be recognized as the Euler-Kockel-Heisenberg correction 
to the effective Lagrangian density, which accounts for the nonlinear effects induced by the external 
field in effective quantum electrodynamics \cite{EU36} - \cite{JE00}. It provides exactly the same 
contribuction obtained when the limit $L \rightarrow \infty$ is considered, i.e., the contribution 
from the boundaries just add a field independent ammount to the E-K-H effective Lagrangian density. 
The independence of both effects clarify the physics governing the behaviour of quantum fields under 
the influence of external fields and/or boundaries conditions. The generalization of the above 
calculation to the twisted case is immediate as well as the inclusion of the fermion mass.
%
%
\section{Concluding Remarks}
Using an approach based on the combination of analytic regularization method throught 
$\alpha$-representation and the Euler-Maclaurin summation formula we had rederived the 
electromagnetic and the fermionic Casimir energy densities. The later, which comprises 
the main results of the present article, was considered in the case of $S^1 \times R^3$ 
space topology, where the role played by the fermion mass and that of an external field 
on the Casimir energy density were fully investigated.

As was shown in section III, the present approach provided a powerfull way to dealwith 
in each step of the calculation, the divergences inherent to the theory. It was found 
that the fermion mass doesn't play any influence on the twisted and untwisted fermionic 
Casimir energy densities, which is in contrast with the first results obtained by Ford 
\cite{FO80}. Experiment may provide the final answer.

We have also seen that, when an external magnetic field is considered, its effect on the 
Casimir energy density appears as an $L$-independent term (which was ultimately identifyed 
with the well known Euler-Kockel-Heisenberg Effective Lagrangian density) plus a term identical 
to that obtained when the external field is absent, expresion (15). This result clearly shows 
the independece of the external field on the boundary conditions, although this seems to be in 
apparent disagreement with the results found in \cite{CO01}.
 
Finally, it must be emphasized that the present construct is a simple and easily generalizable 
method to reexamine many other phenomena. Among these are those related to the Effective Quantum 
Electrodynamics in the context of the ``old fashioned'' Weisskopf's method \cite{WE36}, recentely 
readdressed \cite{JE00}.

%
\vspace{0.9cm} 
\noindent {\bf Acknowledgements.} 
JLT acknowledges IFT/UNESP for the hospitality. LCC is grateful to FAPESP for the financial support.
%

%
}}
\end{document}